# EXTENDED GAUGE MODELS AND $e^-e^- \to W^-W^-$ AT THE NLC

THOMAS G. RIZZO

*Stanford Linear Accelerator Center, Stanford University, Stanford, CA 94309, USA*

## ABSTRACT

We briefly discuss two possible manifestations of the lepton number violating reaction $e^-e^- \to W_i^- W_j^{(*)-}$, which probes the masses and mixings of heavy Majorana neutrinos, at the Next Linear Collider(NLC). Cross sections for this process are shown to be potentially quite large at center of mass energies of order 1-1.5 TeV.

The possibility that some version of the generic process $e^-e^- \to W_i^- W_j^{(*)-}$ might be observable at high energy colliders, such as the NLC, has been a subject of much discussion for almost 15 years[1]. From this choice of notation we want to make it clear from the beginning that this generic reaction represents a rather large number of distinct processes depending on the model under consideration, the most obvious being the production of an on-shell pair of 'left-handed' Standard Model(SM) $W$'s within the $SU(2)_L \otimes U(1)_Y$ framework. Another class of reactions is based on the Left-Right Symmetric Model.

Of course, no matter what the specific process is, a number of model independent requirements must be satisfied. First, since we are dealing with the weakly interacting sector of a renormalizable gauge theory, we require that partial wave unitarity be satisfied for center of mass energies much larger than all other mass scales in the calculation. This is relatively straightforward to enforce when constructing various gauge theory models but some care is still required. Secondly, and just as important, we must remember that the reaction $e^-e^- \to W_i^- W_j^{(*)-}$ is lepton number violating, *i.e.*, it corresponds to a $|\Delta L| = 2$ process in complete analogy with neutrinoless double beta decay, $\beta\beta_{0\nu}$. (In fact, $e^-e^- \to W_i^- W_j^{(*)-}$ is often referred to as *inverse* neutrinoless double beta decay.) This means that whatever specific model is being considered must not only have a mechanism to induce $|\Delta L| = 2$ reactions but that their form and magnitude must be consistent with existing experimental constraints[2] arising from $\beta\beta_{0\nu}$. The most common way of introducing $|\Delta L| = 2$ operators is through Majorana mass terms in the neutrino mass matrix which requires an extension to (at least) the fermion sector of the SM. $e^-e^- \to W^-W^-$ then takes place via $t$- and $u$-channel Majorana neutrino exchanges where the couplings of the neutrinos to the $W$ and electron are model dependent. Clearly, in the limit of vanishing Majorana neutrino masses, the $e^-e^- \to W^-W^-$ amplitude vanishes.

It is easy to see why the simplest two neutrino scheme of this kind which includes the see-saw mass generation pattern cannot yield observable cross sections at the NLC.

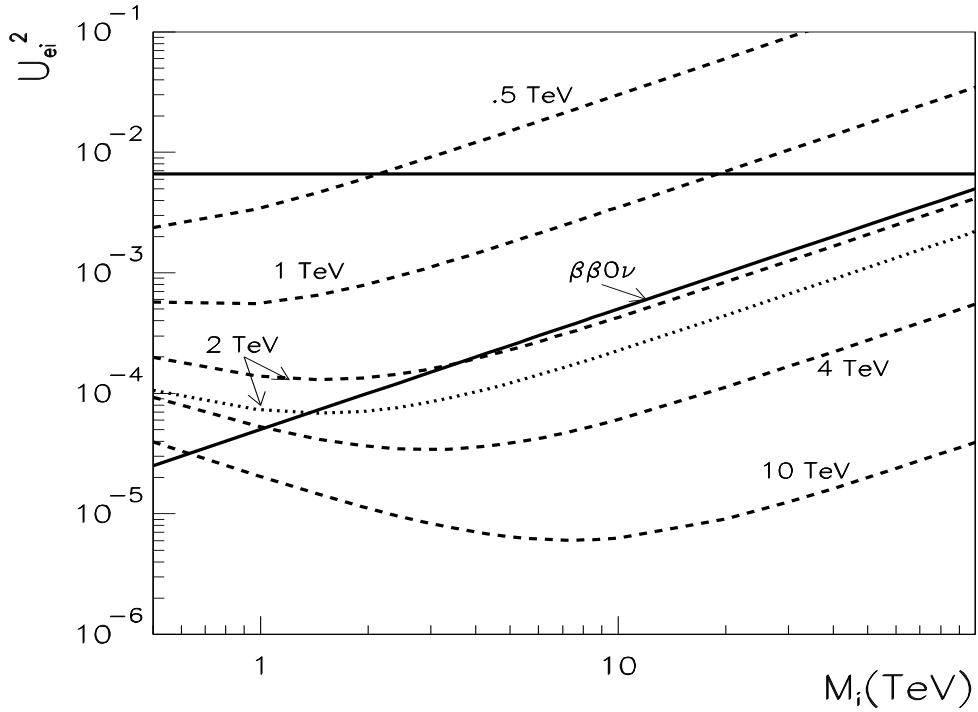

Fig. 1. Discovery limits for $e^-e^- \to W^-W^-$ at the NLC with various center of mass energies(dashed lines). $M_i$ is the heavy neutrino mass while $U_{ei}^2$ is the square of the relevant mixing angle. Unpolarized beams are assumed together with a luminosity of $80(s/1\ TeV^2)fb^{-1}$. The dotted curve in the 2 TeV case corresponds to polarized beams. The region being probed is above the curves in all cases and 10 events are required for discovery. The diagonal solid line is the $\beta\beta_{0\nu}$ constraint of Bélanger *et al.* while the horizontal solid line arises from universality. Here, the allowed region is below the line.

If $D$ and $M$ represent the Dirac and Majorana terms in the 2x2 mass matrix then, in the limit $D/M \ll 1$, the two eigenvalues are $m_1 \simeq -D^2/M$ and $m_2 \simeq M$ with a mixing angle $\phi \simeq -D/M$. $\phi$ is thus forced to be small by the same mechanism which induces the hierarchy in the see-saw mass relation. The amplitude for the $i^{th}$ neutrino exchange behaves as $A_i \sim m_i F(m_i^2/s, M_W^2/s)$, with $F \to 1$ in the limit that $s \to \infty$, thus vanishing as $m_i \to 0$. If we sum over both neutrino exchanges and use the mass and mixing relations above, we find that the *total* amplitude behaves in the $s \to \infty$ limit as $A \sim M[F(m_2) - F(m_1)]\phi^2$. Here, unitarity is clearly satisfied as the difference in the brackets vanishes in the high energy limit. We see, however, that the cross section is now proportional to $\phi^4$ and we recall that $\phi$ must be small. To see how devastating this mixing angle suppression is, let us assume that $M = 500$ GeV, $\phi = 10^{-2}$ and $\sqrt{s} = 1$ TeV. A complete calculation then yields a cross section of $\simeq 0.015$ fb, which is far too small to be observable at the NLC with an integrated luminosity of 100 $fb^{-1}$.

Clearly, within the SM framework, the only way around this disappointing result is to add additional massive neutral fermion fields and to search for a fine-tuning solution to maximize the cross section (*i.e.*, up to the few fb level) while trying to satisfy all the other constraints. As mentioned above, the $\beta\beta_{0\nu}$ constraint is particularly important in the case of heavy neutrino exchange and is a source of controversy amongst the various groups who disagree about the size of the uncertainty in the nuclear matrix elements. Bélanger *et al.*[1] argue that these uncertainties are no more than a factor of 2-3 while Heusch and Minkowski[1] claim that they are substantially larger. The impact of this one constraint can be crucial as shown in Fig.1 from the work of Bélanger *et al.*.. We see that a 2 TeV NLC with polarized beams will be barely able to produce an observable rate and that generally higher energy machines are required. Heusch and Minkowski argue that the $\beta\beta_{0\nu}$ bound is set far too low in this plot and that a 1 TeV machine is sufficient to obtain a reasonable cross section. It is clear that a thorough examination of the disputed nuclear matrix elements is in order so that this situation can be rectified in the near future.

If we go beyond the SM gauge group, the $W$'s need not be associated with the $SU(2)_L$ group factor. The quintessential example of this situation is provided by the Left-Right Symmetric Model[3] based on the gauge group $SU(2)_L \otimes SU(2)_R \otimes U(1)$. In this model we can imagine not just the SM $W^-W^-$ production process but also $W^-W_R^-$ and $W_R^-W_R^-$ as well. But aren't $W_R$'s too massive to be pair produced at the NLC? There are many published lower bounds on the $W_R$ mass, $M_R$, from a number of different sources but *all* of them are easily avoided by the great parameter freedom of the LRM. In fact, the LRM owes much of its survival over the last two decades to the plethora of free 'parameters' it contains: (*a*) the ratio of the $SU(2)_R$ and $SU(2)_L$ gauge couplings, $0.55 \leq \kappa = g_R/g_L \leq 2$, with the lower limit being forced upon us by the internal consistency of the model and the upper bound is from naturalness arguments; (*b*) the masses of the right-handed(RH) neutrinos, (*c*) the elements of the RH CKM mixing matrix, $V_R$, which are *a priori* different than the conventional $V_L$ of the SM, and (*d*) the existence of other additional particles, *e.g.*, charged Higgs, in the spectrum which may participate in loop graphs for rare processes and may also appear as on-shell final states in $W_R$ decay.

Before we turn to the production of $W_R$'s in $e^-e^-$ collisions we first must justify the possibility that they can indeed be light. So let us briefly discuss how the conventional limits are evaded. First, in this model the magnitude of the mass($M_N$) of the right-handed neutrino, $\nu_R$, is *a priori* unknown. If the usual see-saw mechanism is invoked, $\nu_R$ is expected to appear as a heavy Majorana neutrino with a mass set by the RH breaking scale. One of the well-known low energy limits on $M_R$ arises from the polarized $\mu$ decay experiments[4] which yield bounds of order 480 GeV. Of course if the $\nu_R$'s are more massive than a few hundred MeV then they cannot be produced as final states in either $K$, $\pi$, or $\mu$ decay thus avoiding this low energy bound in a rather trivial way. Most analyses of the LRM assume that the elements of $V_R$ and the conventional CKM matrix, $V_L$, differ at most by phase factors. If $V_R = V_L$, then it has been known for some time that considerations of the $K_L - K_S$ mass difference result in a rather strong lower bound[5,6] on the mass of $W_R$ of order 1.6 TeV(or more), thus placing it outside the search capabilities of existing colliders as well as the NLC. However, if we remove the constraint of $V_R = V_L$ and allow $V_R$ to be arbitrary, even in the absence of fine-tuning, we find that $W_R$ can be as light as 280 GeV for a top quark mass of 180 GeV. (Recall that $V_R$ has nine free parameters: three mixing angles and six phases.)

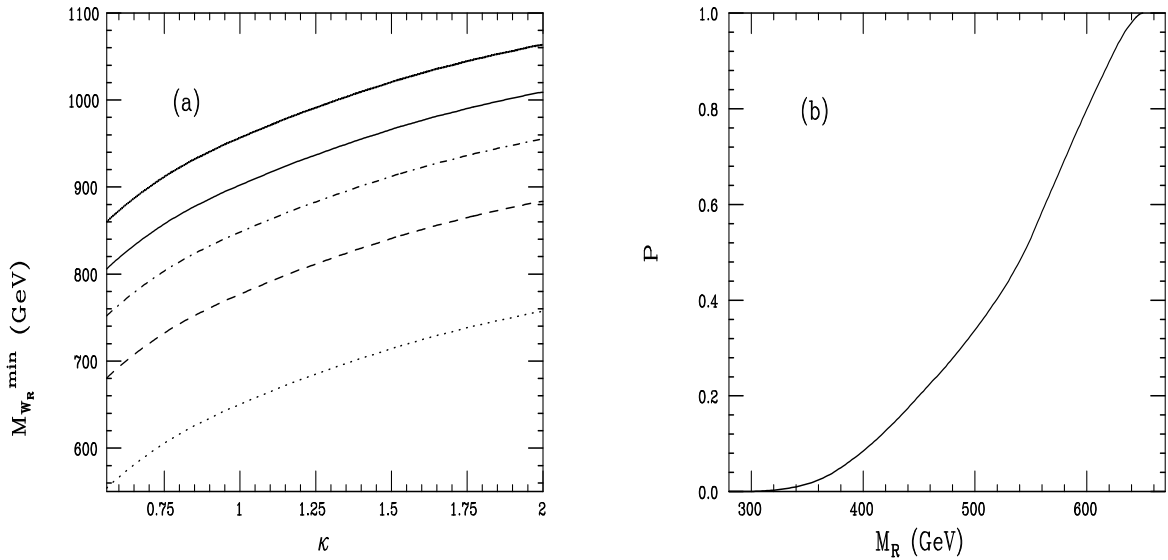

Fig. 2. (a)Tevatron $W_R$ search reach as a function of $\kappa_{eff}$ as described in the text for the luminosities of (from bottom to top) Run Ia, 100, 250, 500, and 1000 $pb^{-1}$. (b) Percentage of the $V_R$ parameter space allowing the mass of $W_R$ below a given value from the CDF run Ia search.

The last well-known constraint on $M_R$ arises from direct searches at the Tevatron. Unlike the corresponding $Z'$ searches at hadron colliders, $W'$ searches via the Drell-Yan process have many subtleties; this is most certainly the case within the LRM. The standard CDF/D0 $W'$ search[7] assumes that the $q'\bar{q}W'$ production vertex has SM strength (*i.e.*, (*i*) $\kappa = 1$ and (*ii*) $|V_{L_{ij}}| = |V_{R_{ij}}|$), that the RH neutrino is (*iii*) 'light' and 'stable', appearing as missing $\not{E}_T$ in the detector, and that the $W_R$'s leptonic

branching fraction($B_l$) is the same as the SM value apart from contributions due to open top(*i.e.*, (*iv*) no exotic decay channels are open). If any of these assumptions are invalid, what happens to the search reach? Assumptions (*i*) and (*iv*) are easily accounted for by the introduction of an effective $\kappa$ parameter, $\kappa_{eff} = \kappa\sqrt{B_l/B_l^{SM}}$ which simply adjusts the overall cross-section normalization with the resulting reach shown in Fig.2a. If assumption (*ii*) is invalid, a significant search reach degradation occurs as is shown in Fig.2b for the CDF run Ia result; *e.g.*, one finds via a Monte Carlo study[8] that for 50(10)% of the $V_R$ parameter space, the CDF run Ia $W_R$ limit is weakened to less than 550(400) GeV. This reduction is a result of modifying the weight of the various parton luminosities which enter into the calculation of the production cross-section. Life gets *much* harder if $\nu_R$ does not appear as missing $E_T$. A massive $\nu_R$ will most likely decay within the detector to $\ell^\pm + jj$, with either charge sign equally likely if $\nu_R$ is a Majorana fermion. D0[7] has looked for $W_R$'s decaying in this manner and has ruled out a reasonable portion of the $M_R - M_N$ plane under the assumption that $\kappa = 1$ and $V_L = V_R$. For $M_N < M_R$, the approximate region $M_R < 550$ GeV and $M_N < 270$ GeV is found to be excluded. Perhaps the worst case scenario is when $\nu_R$ is more massive than $W_R$ so that $W_R$ has only hadronic (or exotic) decay channels open. Can $W_R$ be seen as a bump in dijets? Clearly the chances are somewhat better at the Tevatron, where $S/B$ is perhaps manageable given reasonable statistics, than at the LHC. CDF has already performed such an analysis with run Ia data[7] with somewhat limited results in that only a narrow $M_R$ range is presently excluded even when $\kappa = 1$ and $V_L = V_R$ is assumed. All of these searches will improve dramatically, however, as the Tevatron luminosity increases.

At the NLC, the amplitude for $e^-e^- \to W_R^- W_R^-$ (or $W_R^- W_R^{*-}$) receives both $t-$ and $u-$channel contributions from the RH neutrino in a manner similar to what we saw in the case of the SM discussion above. Apart from a factor of $\kappa$ this coupling strength is unity and no small mixing angles are involved. However, there is also a new $s-$channel contribution from the exchange of a doubly-charged Higgs boson($\Delta$), with mass $M_\Delta$. This doubly-charged Higgs field is a natural consequence of both $SU(2)_R$ breaking as well as the generation of the Majorana mass term for the neutrino and the see-saw mechanism. Since the $e^-e^-\Delta$ coupling is proportional to $M_N$ and the $e^-NW_R$ coupling is chiral, the total amplitude is found to be proportional to $M_N$. The combination of exchanges in all three channels preserves unitarity and can be easily made to satisfy the $\beta\beta_{0\nu}$ constraints since $W_R$ is so massive and we have complete freedom in the $V_R$ elements. At NLC energies, *i.e.*, $\sqrt{s} = 0.5 - 1.5$ TeV, the cross section for $e^-e^- \to W_R^- W_R^-$ is quite large (of order a few $pb$) provided it is kinematically accessible. The production rates are fairly sensitive to the values of $M_N$ and $M_\Delta$, and the cross section is found to have a rather flat angular distribution due to the dominance of the longitudinal-longitudinal helicity amplitude. Of particular interest is the possibility that the $s-$channel $\Delta$ may appear as a resonance depending upon the value of $\sqrt{s}$.

Unfortunately, the 'reach' of the NLC for $W_R$ pair production is rather limited since we are constrained to $W_R$ masses less than $\sqrt{s}/2$. It is thus reasonable to contemplate that like-sign $W_R$ pair production may not be kinematically accessible at these center of mass energies. This forces us to consider the possibility of *singly* producing $W_R$'s via the reaction $e^-e^- \to W_R^-(W_R^-)^* \to W_R^- jj$. We limit ourselves to this $jj$ mode to allow for the possibility that $M_N > M_R$ in which case $W_R$ can only decay to $jj$ (barring the existence of exotic decay modes). Of course, allowing one of the $W_R$'s to be off-shell we are forced to pay the price of an additional gauge coupling as well as three-body phase space. This results in a substantial reduction in the cross section from the on-shell case to the level of a few $fb$. This implies that machine luminosities in the range of $\mathcal{L} = 100 - 200 fb^{-1}$ are required to make use of this channel. The total event rates for $e^-e^- \to W_R^-(W_R^-)^* \to W_R^- jj$ are found in Figs. 4 and 5, in which we have set $\kappa = 1$ and have scaled by an integrated luminosity of $100 fb^{-1}$. Here we see that the reach for $W_R$'s in this channel can be as large as $M_R \simeq 0.8\sqrt{s}$.

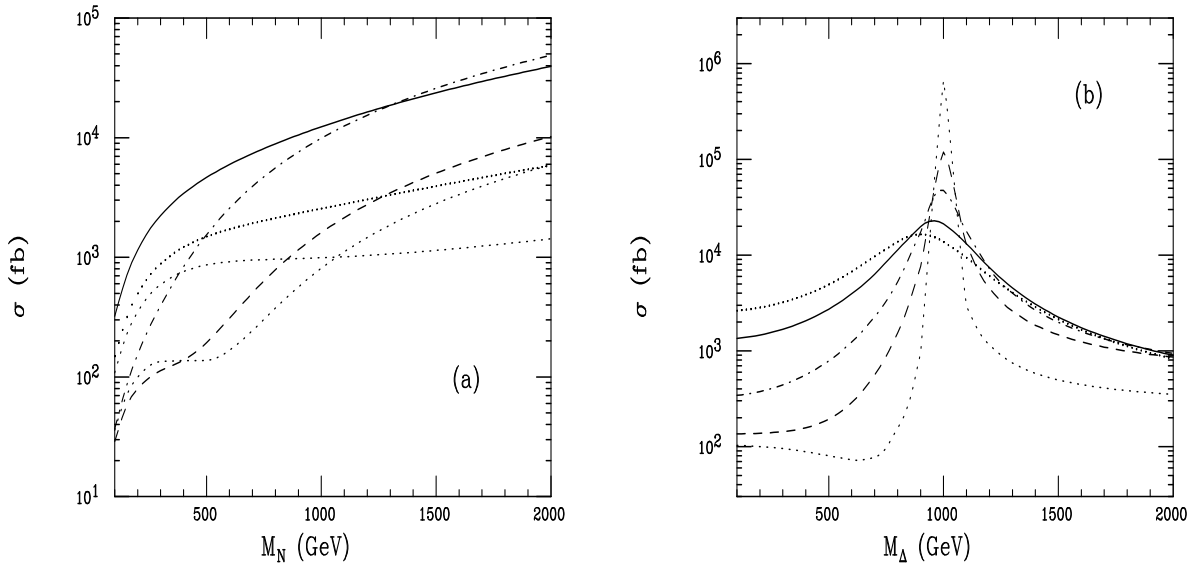

Fig. 3. Cross section for like sign $W_R$ production at a 1 TeV NLC assuming $\kappa = 1$ and $M_R = 480$ GeV as a function of (a)$M_N$ and (b)$M_\Delta$. In(a)[(b)], the curves on the right(left)-hand side correspond, from top to bottom, to $M_\Delta$=800,1200,500,1500,200, and 2000 GeV [$M_N$=1500,1200,800, 500, 200 GeV].

Fig. 4a(b) shows the number of expected $W_R + jj$ events, as a function of $M_R$, at a $\sqrt{s} = 1(1.5)$ TeV $e^-e^-$ collider for different choices of $M_N$ and $M_\Delta$. The results are seen to be quite sensitive to the values of these mass parameters even when $M_R$ is fixed. Setting $M_R = 1$ TeV in Figs. 5a and 5b, we again see reasonable event rates for most choices of $M_N$ and $M_\Delta$ assuming $\sqrt{s} = 1.5$ TeV. The exact rate is, however, a sensitive probe of both the $N$ and $\Delta$ masses. For most choices of the input masses we obtain extremely flat distributions in $cos\theta$ so that detector acceptance is not a problem. However, when $N$ is light a significant angular dependence is observed. This is simply a result of the $t-$ and $u-$ channel poles which develop as $M_N$ tends to zero. Of course, small $M_N$ also leads to a small cross section since the matrix element vanishes in this massless limit.

Potential backgrounds to the process $e^-e^- \to W_R^-(W_R^-)^* \to W_R^- + jj$ at the NLC

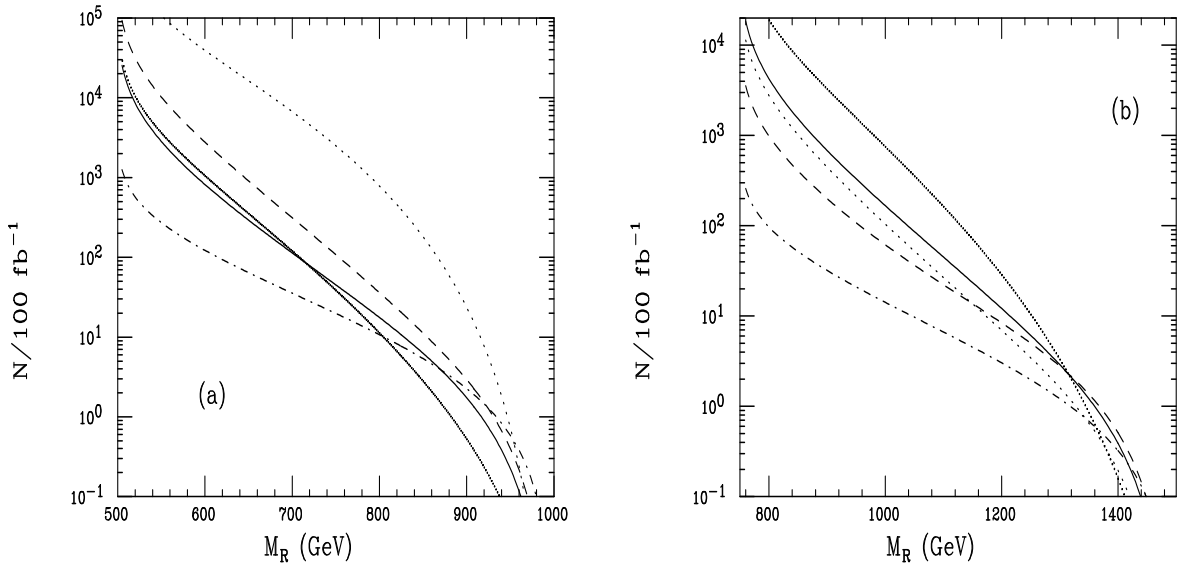

Fig. 4. Event rates per $100fb^{-1}$ for $W_R + jj$ production at a 1 TeV $e^-e^-$ collider assuming $\kappa = 1$ (a) as a function of $M_R$ for $M_N = M_\Delta = 1$ TeV (dots), $M_\Delta = 1.2$ TeV and $M_N = 0.4$ TeV (dashes), $M_\Delta = 0.3$ and $M_N = 0.1$ TeV (dash-dots), $M_\Delta = 2$, $M_N = 0.6$ TeV (solid), or $M_\Delta = 1.8$ and $M_N = 0.6$ TeV (square dots). (b)Same as (a) but for a 1.5 TeV collider.

are easily controlled and/or removed. For example, there may be some contamination from the SM process $e^-e^- \rightarrow W_L^- W_L^- \nu\nu$, but this can be easily eliminated by using missing energy cuts and demanding that the $W_R$ final state be reconstructed from either the $jj$ or $eN \rightarrow eejj$ decay modes. (Since the on-shell $W_R$ decays to either $jj$ or $eN \rightarrow eejj$ there is no missing energy in the signal process.) In addition, with polarized beams, we can take advantage of the fact that $W_R$ couples via right-handed currents while any SM background must arise only via left-handed currents. Within the LRM itself a possible background could arise from a similar lepton-number conserving processes such as $e^-e^- \rightarrow W_R^- W_R^- NN$. Even if such a final state could be produced, in comparison to the process we are considering, the subsequent $N$ decays would lead to a final state with too many charged leptons and/or jets.

In summary, the process $e^-e^- \rightarrow W_i^- W_j^{(*)-}$ offers an interesting opportunity to explore the masses and mixings of heavy Majorana neutrinos at the NLC. In the case where a pair of conventional left-handed $W$'s is produced there remains some controversy as to whether large cross sections can be made consistent with the experimental bounds from $\beta\beta_{0\nu}$. This issue must be cleared up soon. We have noted that there are a number of existing constraints which point to $W_R$ being heavy but that they can be evaded by taking full advantage of the parameter freedom of the LRM. In the LRM, it is also easy to satisfy $\beta\beta_{0\nu}$ bounds due to the flexibility of the right-handed CKM mixing matrix and the large value of the $W_R$ mass with the result that very large cross sections are obtained at the NLC up to the kinematic limit. To go to larger masses, we have considered the production of one on-shell, one off-shell $W_R$ with the off-shell $W_R^* \to jj$. We pay a big price to do this, *i.e.*, an additional power of the coupling constant as well as three-body phase space, but the resulting rates are still significant out to $M_R \simeq 0.8\sqrt{s}$ for integrated luminosities in excess of $100~fb^{-1}$.

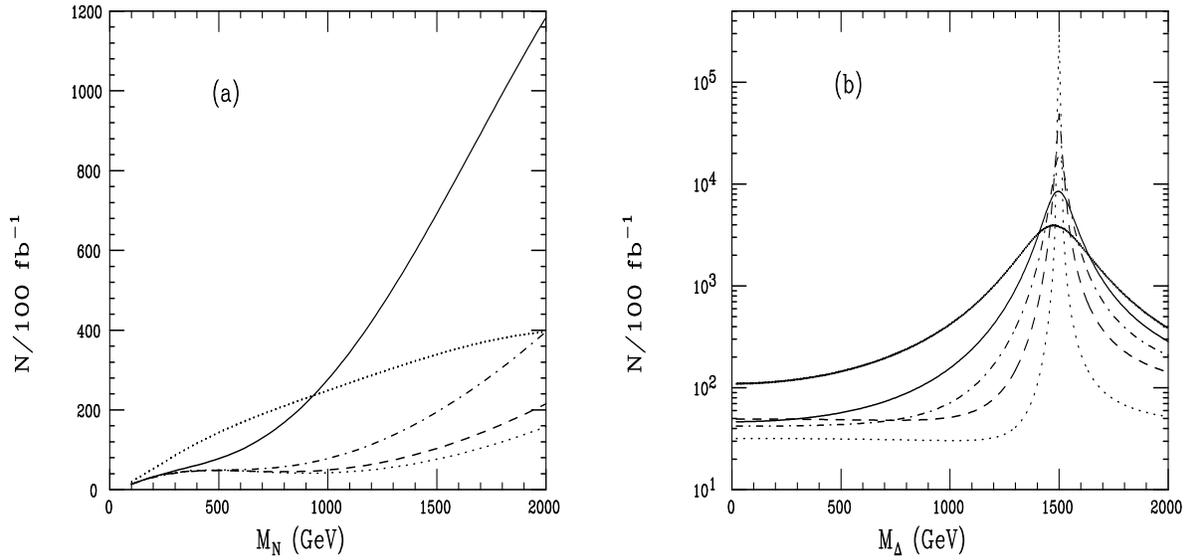

Fig. 5. Event rates per 100 $fb^{-1}$ for $W_R + jj$ production at a 1.5 TeV $e^-e^-$ collider assuming $\kappa = 1$ and $M_R$=1 TeV (a)as a function of $M_N$ for $M_\Delta$=0.3(0.6,1.2,1.5,2) TeV corresponding to the dotted(dashed, dash-dotted, solid, square-dotted) curve; (b)as a function of $M_\Delta$ for $M_N$=0.2(0.5,0.8,1.2,1.5) TeV corresponding to the dotted(dashed, dash-dotted, solid, square-dotted) curve.